\documentclass[journal]{IEEEtran}

\ifCLASSINFOpdf
\else
   \usepackage[dvips]{graphicx}
\fi
\usepackage{url}

\usepackage{graphicx}
\usepackage{booktabs}
\usepackage{amsmath,amsfonts}
\usepackage{physics}
\usepackage{array}
\usepackage[colorlinks = true]{hyperref}

\usepackage{xcolor}
\definecolor{BrewerSetRed}{RGB}{228,26,28}
\definecolor{BrewerSetBlue}{RGB}{55,126,184}
\definecolor{BrewerSetGreen}{RGB}{77,175,74}
\definecolor{BrewerSetPurple}{RGB}{152,78,163}
\definecolor{BrewerSetOrange}{RGB}{255,127,0}
\definecolor{BrewerSetYellow}{RGB}{236,176,23}
\definecolor{BrewerSetBrown}{RGB}{166,86,40}
\definecolor{BrewerSetPink}{RGB}{247,129,191}
\definecolor{BrewerSetGray}{RGB}{153,153,153}

\usepackage{tikz}
\usetikzlibrary{shapes.geometric,positioning}
\DeclareRobustCommand\SolidLine[1]{
    \tikz[baseline=-0.6ex]\draw[very thick, #1] (0,0) -- (2em,0);
}

\begin{document}

\title{Generating Localized Audible Zones Using a Single-Channel Parametric Loudspeaker}

\author{Tao Zhuang, \IEEEmembership{Member, IEEE}, Shaozhe Li, Feng Niu, Jia-Xin Zhong, \IEEEmembership{Member, IEEE}, and Jing Lu, \IEEEmembership{Member, IEEE}
\thanks{This work is supported by National Natural Science Foundation of China (12274221) and the AI \& AI for Science Project of Nanjing University. 
The associate editor coordinating the review of this manuscript and approving it for publication was xxx. (Corresponding author: Jia-Xin Zhong.)}
\thanks{Tao Zhuang, Shaozhe Li and Jing Lu are with the Key Laboratory of Modern Acoustics, Nanjing University, Nanjing 210093, China, also with the NJU-Horizon Intelligent Audio Lab, Horizon Robotics, Beijing 100094, China (e-mail: taozhuang@smail.nju.edu.cn, shaozhe.li@smail.nju.edu.cn and lujing@nju.edu.cn).}
\thanks{Feng Niu is with the Division of Mechanics and Acoustics, National Institute of Metrology, Beijing 100029, China (e-mail: niufeng@nim.ac.cn).}
\thanks{Jia-Xin Zhong is with the Graduate Program in Acoustics, The Pennsylvania State University, University Park, PA 16802, USA (e-mail: jiaxin.zhong@psu.edu).}}

\markboth{Journal of \LaTeX\ Class Files, Vol. 14, No. 8, August 2015}
{Shell \MakeLowercase{\textit{et al.}}: Bare Demo of IEEEtran.cls for IEEE Journals}
\maketitle

\begin{abstract}

Advanced sound zone control (SZC) techniques typically rely on massive multi-channel loudspeaker arrays to create high-contrast personal sound zones, making single-loudspeaker SZC seem impossible. 
In this Letter, we challenge this paradigm by introducing the multi-carrier parametric loudspeaker (MCPL), which enables SZC using only a single loudspeaker. 
In our approach, distinct audio signals are modulated onto separate ultrasonic carrier waves at different frequencies and combined into a single composite signal. 
This signal is emitted by a single-channel ultrasonic transducer, and through nonlinear demodulation in air, the audio signals interact to virtually form multi-channel outputs. 
This novel capability allows the application of existing SZC algorithms originally designed for multi-channel loudspeaker arrays.
Simulations validate the effectiveness of our proposed single-channel MCPL, demonstrating its potential as a promising alternative to traditional multi-loudspeaker systems for achieving high-contrast SZC.
Our work opens new avenues for simplifying SZC systems without compromising performance.

\end{abstract}


\IEEEpeerreviewmaketitle

\section{Introduction}


\IEEEPARstart{S}{ound} zone control (SZC) aims at adjusting the output of a loudspeaker array to brighten or darken a target zone \cite{Druyvesteyn1997, Wu2011, Betlehem2015, Ahrens2012}. 
This technology has applications in environments such as aircraft and/or car cabins \cite{Widmark2019, So2019}, as well as virtual reality (VR) systems \cite{Murphy2011, Zhang2017}. 
However, implementing SZC using electro-dynamic loudspeaker (EDL) arrays typically requires a large array size \cite{Zhang2017, Choi2002, Donley2018}.
Compared to EDL arrays, parametric loudspeaker (PL) arrays offer advantages such as smaller size and flexible phase control \cite{Shinagawa2008, Shi2014, Hahn2021, Zhong2022a}.
PL arrays have been widely used in various audio applications, such as spatial sound reproduction \cite{Tan2012}, virtual sound source construction \cite{Harma2008}, stereo reproduction \cite{Aoki2012}, and acoustic measurements \cite{Arnela2025}, owing to their ability to generate highly directional audio beams \cite{Gan2012, Shi2015, Zhong2024}.
Despite their sharp directivity, the audio beams produced by PLs attenuate slowly along the propagation direction. 
This characteristic makes PLs poorly suited for reverberant environments, such as meeting or listening rooms, due to strong reflections from surrounding walls \cite{Zhong2023a} and scattering from human heads \cite{Zhong2022c}.
The use of multi-channel PLs can mitigate this issue by simultaneously controlling both the directivity and propagation distance of the audio beams \cite{Zhong2022a}. However, both multi-channel EDL arrays and multi-channel PL arrays face practical challenges, as they require a large number of digital-to-analog converters (DACs) and extensive multi-channel wideband signal processing when the channel number is large.

The physical mechanism of PLs relies on modulating an audio signal onto an ultrasonic carrier wave.
This modulated signal is emitted by ultrasonic transducers and subsequently self-demodulated in air, reproducing the audio signal \cite{Gan2012, Zhong2024}. 
Unlike EDLs, it is interesting to note that PL systems offer an additional approach for wave manipulation through the ultrasonic carrier wave, a property that has been largely underexplored.
In this Letter, we proposes the concept of a \emph{single-channel multi-carrier parametric loudspeaker (MCPL)}. 
In our approach, distinct audio signals are modulated onto separate ultrasonic carrier waves at different frequencies and combined into a single composite signal. 
This signal is emitted by a single-channel ultrasonic transducer, and through nonlinear self-demodulation in air, the audio signals interact to virtually form multi-channel outputs.
Previous studies have explored dual-carrier ultrasound PLs to create length-limited audio beams by canceling the audio sound pressure at a far-field point along the axis, thereby controlling the audio propagation distance \cite{Hedberg2010, Zhu2023, Nomura2012, Nomura2025}. 
This method can be regarded as a special case of our proposed MCPL when the carrier number is limited to two.

The single-channel MCPL we propose generalizes this concept by utilizing $N$ carriers, where the frequency difference between any two carriers exceeds 20\,kHz. 
This condition ensures that the audio signals demodulated from different carriers can be linearly superposed.
More importantly, this treatment allows us to apply the SZC algorithms---originally designed for conventional EDL arrays \cite{Druyvesteyn1997, Widmark2019, Choi2002} in linear acoustics---to the single-channel MCPL, despite its inherently nonlinear nature.
By employing the acoustic contrast control (ACC) algorithm, we optimize the weight coefficients of each carrier channel to maximize the contrast between near-field and far-field audio sound, thereby achieving better localized sound reproduction.
Notably, the proposed method modulates the same audio signal onto multiple ultrasonic carriers, applies appropriate weighting, and sums the resulting signals before outputting them to the PL. 
This approach requires only a single-channel signal for playback, meaning that only one DAC is needed.
This significantly simplifies hardware implementation compared to conventional multi-channel PL or EDL arrays.
The performance of the single-channel MCPL is evaluated and compared to conventional methods, highlighting its advantages in achieving precise and flexible sound reproduction.

\section{\label{sec:Theory} Theory}

\subsection{Audio sound generated by the MCPL}
\label{sec:audio_press_PL_array}

\begin{figure}[!t]
\centering
\includegraphics[width = 0.38\textwidth]{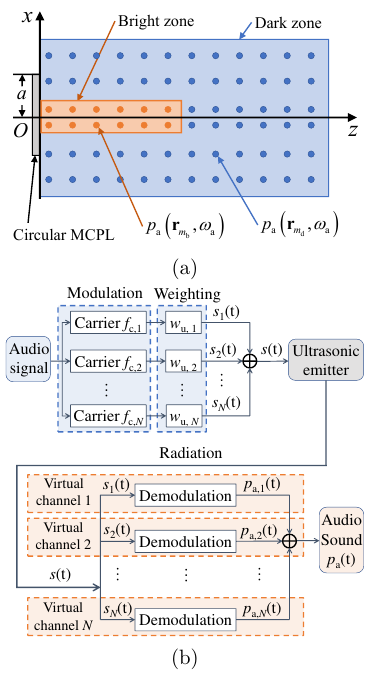}
\vspace{-1em}
\caption{
\label{fig:sketch_MCPLArray}
(a) Sketch of a circular MCPL positioned on the $Oxy$-plane for generating bright and dark zones. 
The $y$-axis is omitted for simplicity. 
(b) Signal flow diagram. 
Audio signals are modulated onto $N$ carrier waves, combined into a single-channel signal $s(t)$, and radiated by an ultrasonic emitter to produce $N$ virtual channels in air.
}
\vspace{-15pt}
\end{figure}

As shown in Fig.\,\ref{fig:sketch_MCPLArray}\,(a), a circular single-channel MCPL with a radius of $a$ is assumed to be located in the plane $Oxy$. 
As shown in Fig.\,\ref{fig:sketch_MCPLArray}\,(b), for a single-channel MCPL proposed in this work, a single-channel audio signal at frequency $f_\mathrm{a}$ is modulated by $N$ distinct ultrasonic carrier wave with frequency $f_{\mathrm{c},n}, (n = 1,2,...,N)$. 
The modulated signal $s_n (t)$ by the ultrasonic carrier wave with frequency $f_{\mathrm{c},n}$ can be expressed as
\begin{equation}
    s_n \qty(t) = w_{1,n} \exp \qty(- \mathrm{i} \omega_{1, n} t) + w_{2,n} \exp \qty(- \mathrm{i} \omega_{2, n} t), 
\end{equation}
where i is the imaginary unit, $w_{\mathrm{u},n}, (\mathrm{u} = 1,2)$ is dimensionless complex number used to control the output amplitude and phase of the lower and upper sideband ultrasound at frequencies $\omega_{1, n} = 2 \pi (f_{\mathrm{c},n} - f_\mathrm{a}/2)$ and $\omega_{2, n} = 2 \pi (f_{\mathrm{c},n} + f_\mathrm{a}/2)$, respectively. 

It is noteworthy that the modulation of $N$ ultrasonic carrier waves and their corresponding weighting can be preprocessed in the digital domain. 
All modulated signals, $s_n (t)$, are then summed to produce a single composite signal, $s(t)$.
This composite signal is converted to an analog format using a DAC and drives a single-channel ultrasonic emitter. 
We refer to this approach as a ``single-channel loudspeaker'' technique because it requires only one physical loudspeaker to reproduce the audio signal, unlike conventional loudspeaker arrays that rely on multiple speakers, each driven by separate signals.

Then the audio sound in air generated by modulated signal $s_n (t)$ can be obtained by the quasilinear solution of the Westervelt equation \cite{Zhong2024, Zhong2021, Li2024}, shown as
\begin{equation}
 p_{\mathrm{a}, n} \qty(\vb{r}, \omega_\mathrm{a}) = w_n H_{\mathrm{a}, n} \qty(\vb{r}, \omega_\mathrm{a}).
\label{eq:west_audio_press}
\end{equation}
Here, $\omega_\mathrm{a} = 2 \pi f_\mathrm{a}$, $w_n = w_{1,n}^* w_{2,n}$ is a dimensionless complex number to control the output amplitude and phase of the audio sound generated by the $n$-th carrier, the superscript ``*'' denotes the complex conjugate transpose, and $H_{\mathrm{a}, n} \qty(\vb{r}, \omega_\mathrm{a})$ represents the transfer function, shown as
\begin{equation}
    H_{\mathrm{a}, n} \qty(\vb{r}, \omega_\mathrm{a}) = - \mathrm{i} \rho_0 \omega_\mathrm{a} \iiint_{-\infty}^\infty {
     \frac{ q_n \qty(\vb{r}_\mathrm{v},\omega_\mathrm{a}) } 
     {4 \pi \abs{\vb{r} - \vb{r}_\mathrm{v}}}
     \mathrm{e}^{\mathrm{i} k_\mathrm{a} \abs{\vb{r} - \vb{r}_\mathrm{v}}}
     \dd[3]{\vb{r}_\mathrm{v}}
 }, 
\end{equation}
where $k_\mathrm{a} = \omega_\mathrm{a} / c_0$ is the wavenumber of 
the audio sound, $\rho_0$ is the air density, $c_0$ is the sound velocity in air and $q \qty(\vb{r}_\mathrm{v},\omega_\mathrm{a})$ is the virtual audio sources with the source density of \cite{Zhong2021}
\begin{equation}
q_n (\vb{r}_\mathrm{v},\omega_\mathrm{a}) 
= - \frac{\mathrm{i} \beta \omega_\mathrm{a}} {\rho_0^2 c_0^4}  
    p_{1, n}^* \qty(\vb{r}_\mathrm{v},\omega_{1, n})
    p_{2, n} \qty(\vb{r}_\mathrm{v},\omega_{2, n}), 
\label{eq:PAL_array_virtual_source}
\end{equation}
where $\beta$ is the nonlinearity coefficient, $p_{1, n} \qty(\vb{r}_\mathrm{v},\omega_{1, n})$ and $p_{2, n} \qty(\vb{r}_\mathrm{v},\omega_{2, n})$ is the ultrasound in air generated by $\exp(- \mathrm{i} \omega_{1, n} t)$ and $\exp(- \mathrm{i} \omega_{2, n} t)$, respectively, which can be obtained by the Rayleigh integral as \cite{Cervenka2013}
\begin{equation}
    p_{\mathrm{u}, n} \qty( \vb{r}_\mathrm{v}, \omega_{\mathrm{u}, n} ) 
    =  - \frac{\mathrm{i} \omega_{\mathrm{u}, n} \rho_0 v_0}{2 \pi} 
    \iint_S 
    \frac{\exp(\mathrm{i} k_{\mathrm{u}, n} d_\mathrm{s})}{d_\mathrm{s}} 
    \mathrm{d}^2 \vb{r}_\mathrm{s}.
 	\label{eq:Ray_ultra_press}
\end{equation}
Here, $v_0$ is the surface vibration velocity of the MCPL, which is set as 1\,m/s here, $k_{\mathrm{u}, n}$ is the complex wavenumber of frequency $\omega_{\mathrm{u}, n}$, $\vb{r}_\mathrm{s}$ represents the surface source point of the MCPL and $d_\mathrm{s} = \abs{\vb{r}_\mathrm{v} - \vb{r}_\mathrm{s}}$ represents the distance between the surface source point $\vb{r}_\mathrm{s}$ and the virtual source point $\vb{r}_\mathrm{v}$.
When the center frequency $f_{\mathrm{c}, n}$ between each carrier exceeds 20\,kHz, the audio sound generated by the MCPL can be expressed as
\begin{equation}
    p_\mathrm{a} \qty(\vb{r}, \omega_\mathrm{a}) = \sum_{n=1}^N p_{\mathrm{a},n} \qty(\vb{r}, \omega_\mathrm{a}) = \sum_{n=1}^N w_n H_{\mathrm{a},n} \qty(\vb{r}, \omega_\mathrm{a}).
    \label{eq:audio_multi-carrier_PL}
\end{equation}
Although the audio sound generated by the MCPL is a nonlinear process, it is noteworthy that the total audio output, as shown in (\ref{eq:audio_multi-carrier_PL}), is simply a linear combination of the contributions from each modulated signal, $s_n(t)$.
As a result, this can be equivalently interpreted as $N$ audio channels being virtually created in the air.
This interpretation enables the application of existing SZC methods, originally developed for conventional EDL arrays, to achieve effective contrast control between specified bright and dark zones.

\subsection{Acoustic contrast control using a single-channel MCPL}

As shown in Fig.\,\ref{fig:sketch_MCPLArray}\,(a), $M_\mathrm{b}$ and $M_\mathrm{d}$ control points are assumed in the bright zone and dark zone, respectively. 
The audio sound generated by the single-channel MCPL in the bright zone and dark zone can be expressed as 
\begin{equation}
    \vb{p}_\mathrm{a,b} = \vb{H}_\mathrm{b} \vb{w}, \vb{p}_\mathrm{a,d} = \vb{H}_\mathrm{d} \vb{w},
    \label{eq:audio_sound_vector_MCPL}
\end{equation}
where $\vb{p}_\mathrm{a,b} = [p_\mathrm{a}(\vb{r}_1) \text{ } p_\mathrm{a}(\vb{r}_2) \cdots p_\mathrm{a}(\vb{r}_{M_\mathrm{b}}) ]^\mathrm{T}$, $\vb{p}_\mathrm{a,d} = [p_\mathrm{a}(\vb{r}_1) \text{ } p_\mathrm{a}(\vb{r}_2) \cdots p_\mathrm{a}(\vb{r}_{M_\mathrm{d}}) ]^\mathrm{T}$ are the audio sound vectors in the bright and dark zones, respectively.
The superscript ``T'' represents the transpose. 
The transfer matrix of the bright zone $\vb{H}_\mathrm{b}$ and the dark zone $\vb{H}_\mathrm{d}$ are expressed as
\begin{equation}
    \vb{H}_\mathrm{b} = 
    \begin{bmatrix}
        H_{\mathrm{a}, 1} (\vb{r}_1) & H_{\mathrm{a}, 2} (\vb{r}_1) & \cdots & H_{\mathrm{a}, N} (\vb{r}_1) \\
        H_{\mathrm{a}, 1} (\vb{r}_2) & H_{\mathrm{a}, 2} (\vb{r}_2) & \cdots & H_{\mathrm{a}, N} (\vb{r}_2) \\
        \vdots & \vdots & \ddots & \vdots \\
        H_{\mathrm{a}, 1} (\vb{r}_{M_\mathrm{b}}) & H_{\mathrm{a}, 2} (\vb{r}_{M_\mathrm{b}}) & \cdots & H_{\mathrm{a}, N} (\vb{r}_{M_\mathrm{b}})
    \end{bmatrix}, 
\end{equation}
\begin{equation}
    \vb{H}_\mathrm{d} = 
    \begin{bmatrix}
        H_{\mathrm{a}, 1} (\vb{r}_1) & H_{\mathrm{a}, 2} (\vb{r}_1) & \cdots & H_{\mathrm{a}, N} (\vb{r}_1) \\
        H_{\mathrm{a}, 1} (\vb{r}_2) & H_{\mathrm{a}, 2} (\vb{r}_2) & \cdots & H_{\mathrm{a}, N} (\vb{r}_2) \\
        \vdots & \vdots & \ddots & \vdots \\
        H_{\mathrm{a}, 1} (\vb{r}_{M_\mathrm{d}}) & H_{\mathrm{a}, 2} (\vb{r}_{M_\mathrm{d}}) & \cdots & H_{\mathrm{a}, N} (\vb{r}_{M_\mathrm{d}})
    \end{bmatrix}, 
\end{equation}
and the weight vector is denoted as $\vb{w} = [w_1 \text{ } w_2 \cdots w_N]^\mathrm{T}$. 
Then the optimization problem using ACC can be formulated as 
\begin{equation}
    \underset{\vb{w}}{\max} \frac{\abs{\vb{p}_\mathrm{a,b}}^2}{\abs{\vb{p}_\mathrm{a,d}}^2},
    \label{eq:ACC}
\end{equation}
where $\abs{\vdot}$ is the norm of the complex vector. 
By substituting (\ref{eq:audio_sound_vector_MCPL}) into (\ref{eq:ACC}), the optimization problem can be expressed as 
\begin{equation}
    \underset{\vb{w}}{\max} \frac{\vb{w}^* \vb{H}_\mathrm{b}^* \vb{H}_\mathrm{b} \vb{w}}{\vb{w}^* \vb{H}_\mathrm{d}^* \vb{H}_\mathrm{d} \vb{w}}.
    \label{eq:optimize_form}
\end{equation}
The optimization problem (\ref{eq:optimize_form}) has the closed solution which is the eigenvector corresponded to the max eigenvalue of the matrix pair $(\vb{H}_\mathrm{b}^* \vb{H}_\mathrm{b}, \vb{H}_\mathrm{d}^* \vb{H}_\mathrm{d})$ \cite{Choi2002}.

\section{\label{sec:Simu} Simulation}

\subsection{Parameters setting}

A circular PL with radius 0.1\,m is considered in this section and the simulations are conducted under free-field conditions. 
The audio sound generated by the proposed single-channel MCPL with different numbers of carriers is simulated in this section. 
For the single-channel MCPL with 1, 2, 3, and 4 carriers, the carrier frequencies are chosen as the first 1, 2, 3, and 4 frequencies from 40 kHz, 80 kHz, 120 kHz, and 160 kHz, respectively.
Notably, the MCPL with a single carrier corresponds to the conventional PL and will serve as a baseline for comparative analysis in this study.
For the proposed single-channel MCPL, the bright zone is set as a rectangular area of $-0.2 \text{ m} \leq x \leq 0.2 \text{ m}$, $0.1 \text{ m} \leq z \leq 1 \text{ m}$ with $10 \times 10$ control points uniformly distributed for calculating the transfer function $\vb{P}_\mathrm{b}$; the dark zone is set as a rectangular area of $-1 \text{ m} \leq x \leq 1 \text{ m}$, $1.5 \text{ m} \leq z \leq 6 \text{ m}$ with the same $30 \times 45$ control points uniformly distributed for calculating the transfer function $\vb{P}_\mathrm{d}$. 

For a fair comparison, the weights of each carrier in different PLs are normalized based on the weight of the carrier with 40\,kHz.
The effective propagation distance of the loudspeaker array can be defined as the position of the farthest point along the $z$-axis where the audio sound pressure level (SPL) has decreased by 10\,dB compared to the maximum on-axis audio SPL.
Other simulation parameters are listed in Table~\ref{tab:para_setting}.
Direct integration to calculate the quasilinear solution of the Westervelt equation requires significant computational effort \cite{Zhong2024}; therefore, in this section, the audio sound field radiated by the PL is calculated using the extended King integral method \cite{Li2024}.

\begin{table}
\caption{\label{tab:para_setting}Parameters used in the simulations}
\label{table}
\small
\setlength{\tabcolsep}{3pt}
\begin{tabular}{|p{120pt}|p{110pt}|}
\hline
Parameters & Value \\  
\hline
Temperature in air & 20~$^\circ$C\\
Humidity in air & 70~\%\\
Air density  & 1.21~kg/$\mathrm{m}^3$\\
Sound speed in air & 343~m/s\\
Nonlinear coefficient & 1.2\\
\hline
\end{tabular}
\label{tab1}
\end{table}

\subsection{\label{sec:results}Results}

\begin{figure}[!t]
\centering
\includegraphics[width = 0.45\textwidth]{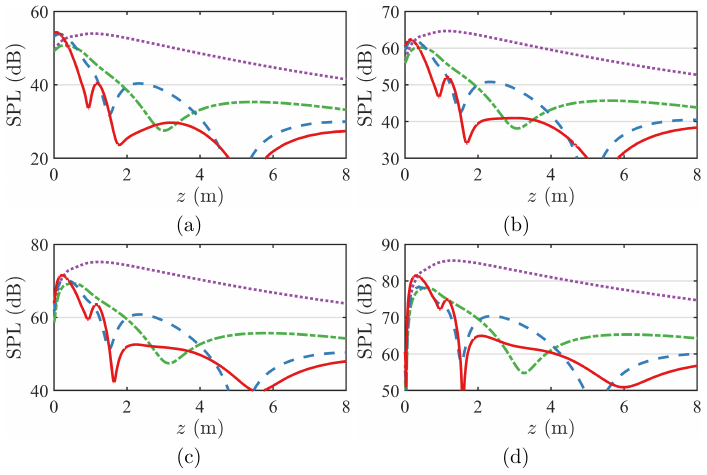}
\vspace{-1em}
\caption{\label{fig:sim_Axial_MCPL}{The axial audio sound generated by \SolidLine{color=BrewerSetPurple, dotted}: the conventional PL  and the proposed single-channel MCPL with \SolidLine{color=BrewerSetBlue, dashed}: 2\,carriers, \SolidLine{color=BrewerSetGreen, dash dot}: 3\,carriers and \SolidLine{color=BrewerSetRed}: 4\,carriers. 
The audio frequency is set as (a) 500\,Hz, (b) 1\,kHz, (c) 2\,kHz and (d) 4\,kHz.}}
\end{figure}

Figure\,\ref{fig:sim_Axial_MCPL} shows the axial audio sound generated by different single-channel MCPLs.
It can be observed that the conventional PL exhibits an excessively long propagation distance, commonly exceeding 8\,m.
This characteristic makes it unsuitable for use in enclosed environments due to strong reflections from surrounding walls \cite{Zhong2023a}.
In contrast, the single-channel MCPL effectively addresses this issue by achieving a well-localized audible zone. 
For example, at an audio frequency of 1\,kHz, the effective propagation distance for the single-channel MCPL with 4\,carriers is approximately 1.8\,m, compared to around 7\,m for the conventional PL. 
Additionally, the propagation distance decreases as the number of carriers increases, indicating improved control performance.
At an audio frequency of 4\,kHz, the effective propagation distance reduces progressively as the number of carriers increases from 1 to 2, 3, and 4, reaching 3.6\,m, 2.5\,m, and 1.5\,m, respectively. 
This behavior arises because the single-channel MCPL utilizes multiple carriers to generate more virtual channels in air, providing greater flexibility in controlling the audio sound distribution and improving the localization of the audible zone.
Furthermore, the size of the localized audible zone generated by the proposed single-channel MCPL can be flexibly controlled by adjusting the bright zone in the ACC method. 


\begin{figure}[!t]
\centering
\includegraphics[width = 0.48\textwidth]{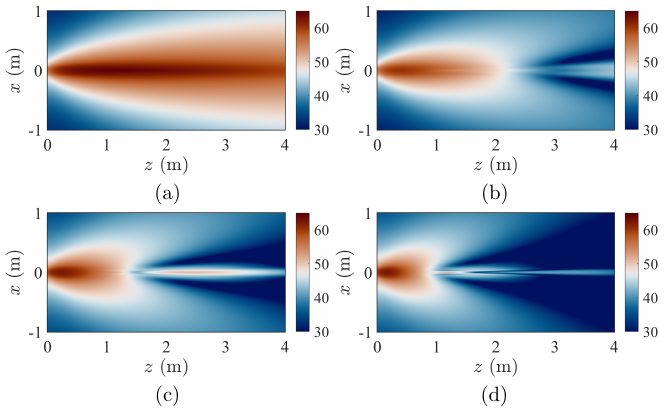}
\vspace{-1em}
\caption{\label{fig:sim_2D_MCPL_diffCarrierNum}{The audio sound distribution on the $Oxz$ plane generated by the conventional PL (a) and the proposed single-channel MCPL with 2 carriers (b), 3 carriers (c) and 4 carriers (d). 
The audio frequency is 1\,kHz}}
\end{figure}

Figure\,\ref{fig:sim_2D_MCPL_diffCarrierNum} shows the audio sound pressure generated by the conventional PL and proposed single-channel MCPLs with different numbers of carriers in the $Oxz$ plane at the same audio frequency. 
It can be observed that, in addition to the phenomenon where the localized sound reproduction of the single-channel MCPL improves with an increasing number of carriers along the axial direction, the localized sound reproduction performance of the single-channel MCPL in the off-axis region also improves as the number of carriers increases.

\begin{figure}[!t]
\centering
\includegraphics[width = 0.48\textwidth]{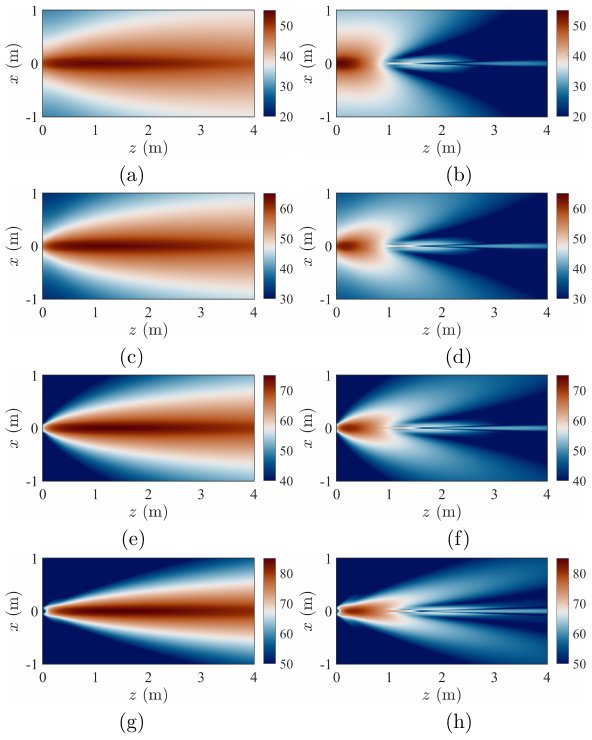}
\vspace{-1em}
\caption{\label{fig:sim_2D_MCPL}{The audio sound distribution on the $Oxz$ plane generated by the conventional PL (left column) and the proposed single-channel MCPL (right column). 
The audio frequency is: (a), (b), 500\,Hz; (c), (d), 1\,kHz; (e), (f), 2\,kHz and (g), (h), 4\,kHz.}}
\end{figure}

Figure\,\ref{fig:sim_2D_MCPL} shows the distribution of audio sound generated by different PLs on the $Oxz$ plane. 
It can be observed that the conventional PL obtains great localized audible zoning in the transverse direction, which is challenging to be obtained by a single-channel EDL.
However, the audio sound generated by the conventional PL shows long propagation distance at all audio frequencies. 
It also can be observed that the single-channel MCPL solves this problem, and achieves better localized audible zone generating compared to the conventional PL not only in the on-axis region, but also in the off-axis region. 
For example, at an audio frequency of 1\,kHz, the single-channel MCPL confines audio sound above 60\,dB in the region approximately $z < 1 \text{ m}$, while the conventional PL confines audio sound above 60\,dB even when $z > 4 \text{ m}$. 
Thus, based on the distribution of audio sound on the $Oxz$ plane, we can also conclude that the single-channel MCPL can generate the localized audible zone effectively.

\section{\label{sec:Conclu} Discussion and Conclusion}

In this work, a single-channel MCPL is proposed and used based on the ACC method to achieve localized audible zones reproduction.
Numerical results validate the feasibility of the proposed method. 
This is because the single-channel MCPL, with its increased number of carriers compared to the conventional PL, can create multiple virtual channels in air, allowing for more effective manipulation of the audio sound field.
Notably, the single-channel MCPL requires multiple carriers modulation only in signal processing; it only needs a single DAC for actual implementation, making it a single-channel system in terms of hardware.
This makes the proposed single-channel MCPL easier to implement in sound reproduction circuitry compared to conventional multi-channel EDL arrays and multi-channel PL arrays.
This work provides a new approach in controlling the audio sound field generated by PLs, and the proposed method enables the effective generation of localized audible zones using a PL on a single-channel basis.


A significant practical challenge of the proposed MCPL system lies in the implementation of wide-band ultrasonic transducers.
In existing literature, PLs are typically constructed using PZT ultrasonic emitters, which operate at a central frequency and have a narrowband response \cite{Zhong2024,Zhu2023}.
Combining transducers with different center frequencies into an array can effectively create a wide-band ultrasonic transducer.
Moreover, recent advancements in micromachined (MEMS) ultrasonic transducers have demonstrated a broad bandwidth compared to PZT transducers \cite{Lee2009, Je2015, Wygant2009, Hwang2016, Ahn2019, Niu2022}, making them a promising alternative for future MCPL fabrication.
This work lays the foundation for developing more flexible audio sound field control and achieving more efficient localized audible zone reproduction.


\bibliographystyle{IEEEbib}
\bibliography{Reference}

\begin{thebibliography}{10}

\bibitem{Druyvesteyn1997}
W.~F. Druyvesteyn and J.~Garas,
\newblock ``Personal sound,''
\newblock {\em J. AUDIO ENG. SOC.}, vol. 45, no. 9, pp. 685--701, 1997.

\bibitem{Wu2011}
Y.~J. Wu and T.~D. Abhayapala,
\newblock ``Spatial multizone soundfield reproduction: Theory and design,''
\newblock {\em IEEE/ACM Trans. Audio, Speech, Lang. Process.}, vol. 19, no. 6,
  pp. 1711--1720, 2011.

\bibitem{Betlehem2015}
T.~Betlehem, W.~Zhang, M.~A. Poletti, and T.~D. Abhayapala,
\newblock ``Personal sound zones: Delivering interface-free audio to multiple
  listeners,''
\newblock {\em IEEE Signal Process. Mag.}, vol. 32, no. 2, pp. 81--91, 2015.

\bibitem{Ahrens2012}
J.~Ahrens,
\newblock {\em Analytic methods of sound field synthesis},
\newblock Springer Science \& Business Media, 2012.

\bibitem{Widmark2019}
S.~Widmark,
\newblock ``Causal mse-optimal filters for personal audio subject to
  constrained contrast,''
\newblock {\em IEEE/ACM Trans. Audio, Speech, Lang. Process.}, vol. 27, no. 5,
  pp. 972--987, May 2019.

\bibitem{So2019}
H.~So and J.-W. Choi,
\newblock ``Subband optimization and filtering technique for practical personal
  audio systems,''
\newblock in {\em ICASSP 2019-2019 IEEE International Conference on Acoustics,
  Speech and Signal Processing (ICASSP)}, May 2019, pp. 8494--8498.

\bibitem{Murphy2011}
D.~Murphy and F.~Neff,
\newblock ``Spatial sound for computer games and virtual reality,''
\newblock in {\em Game sound technology and player interaction: Concepts and
  developments}, pp. 287--312. IGI Global, 2011.

\bibitem{Zhang2017}
W.~Zhang, P.~N. Samarasinghe, H.~Chen, and T.~D. Abhayapala,
\newblock ``Surround by sound: A review of spatial audio recording and
  reproduction,''
\newblock {\em Appl. Sci.}, vol. 7, no. 5, pp. 532, 2017.

\bibitem{Choi2002}
J.-W. Choi and Y.-H. Kim,
\newblock ``Generation of an acoustically bright zone with an illuminated
  region using multiple sources,''
\newblock {\em J. Acoust. Soc. Am.}, vol. 111, no. 4, pp. 1695--1700, 2002.

\bibitem{Donley2018}
J.~Donley, C.~Ritz, and W.~B. Kleijn,
\newblock ``Multizone soundfield reproduction with privacy-and quality-based
  speech masking filters,''
\newblock {\em IEEE/ACM Trans. Audio, Speech, Lang. Process.}, vol. 26, no. 6,
  pp. 1041--1055, 2018.

\bibitem{Shinagawa2008}
K.~Shinagawa, Y.~Ohtomo, H.~Takemura, and H.~Mizoguchi,
\newblock ``Simultaneous generation of multiple three-dimensional sound spot by
  using 512 ch panel loudspeaker array,''
\newblock in {\em 2008 SICE Annual Conference}, 2008, pp. 179--182.

\bibitem{Shi2014}
C.~Shi, Y.~Kajikawa, and W.-S. Gan,
\newblock ``An overview of directivity control methods of the parametric array
  loudspeaker,''
\newblock {\em APSIPA Trans. Signal Inf. Process.}, vol. 3, pp. e20, 2014.

\bibitem{Hahn2021}
N.~Hahn, J.~Ahrens, and C.~Andersson,
\newblock ``Parametric array using amplitude modulated pulse trains:
  Experimental evaluation of beamforming and single sideband modulation,''
\newblock in {\em Audio Engineering Society Convention 151}. 2021, Audio
  Engineering Society.

\bibitem{Zhong2022a}
J.~Zhong, T.~Zhuang, R.~Kirby, M.~Karimi, X.~Qiu, H.~Zou, and J.~Lu,
\newblock ``Low frequency audio sound field generated by a focusing parametric
  array loudspeaker,''
\newblock {\em IEEE/ACM Trans. Audio, Speech, Lang. Process.}, vol. 30, pp.
  3098--3109, 2022.

\bibitem{Tan2012}
E.-L. Tan, W.-S. Gan, and C.-H. Chen,
\newblock ``Spatial sound reproduction using conventional and parametric
  loudspeakers,''
\newblock in {\em Proceedings of The 2012 Asia Pacific Signal and Information
  Processing Association Annual Summit and Conference}, Dec. 2012, pp. 1--9.

\bibitem{Harma2008}
A.~H{\"a}rm{\"a}, S.~Van De~Par, and W.~De~Bruijn,
\newblock ``On the use of directional loudspeakers to create a sound source
  close to the listener,''
\newblock in {\em 124th Audio Engineering Society Convention 2008}, May 2008,
  pp. 160--167.

\bibitem{Aoki2012}
S.~Aoki, M.~Toba, and N.~Tsujita,
\newblock ``Sound localization of stereo reproduction with parametric
  loudspeakers,''
\newblock {\em Appl. Acoust.}, vol. 73, no. 12, pp. 1289--1295, Dec. 2012.

\bibitem{Arnela2025}
M.~Arnela, R.~Burbano-Escol{\`a}, R.~S. Ribeiro, and O.~Guasch,
\newblock ``Reverberation time and random-incidence sound absorption measured
  in the audible and ultrasonic ranges with an omnidirectional parametric
  loudspeaker,''
\newblock {\em Appl. Acoust.}, vol. 229, pp. 110414, 2025.

\bibitem{Gan2012}
W.-S. Gan, J.~Yang, and T.~Kamakura,
\newblock ``A review of parametric acoustic array in air,''
\newblock {\em Appl. Acoust.}, vol. 73, no. 12, pp. 1211--1219, 2012.

\bibitem{Shi2015}
C.~Shi and Y.~Kajikawa,
\newblock ``A convolution model for computing the far-field directivity of a
  parametric loudspeaker array,''
\newblock {\em J. Acoust. Soc. Am.}, vol. 137, no. 2, pp. 777--784, 2015.

\bibitem{Zhong2024}
J.~Zhong and X.~Qiu,
\newblock {\em Acoustic Waves Generated by Parametric Array Loudspeakers},
\newblock CRC Press, 2024.

\bibitem{Zhong2023a}
J.~Zhong, H.~Zou, and J.~Lu,
\newblock ``A modal expansion method for simulating reverberant sound fields
  generated by a directional source in a rectangular enclosure,''
\newblock {\em J. Acoust. Soc. Am.}, vol. 154, no. 1, pp. 203--216, 2023.

\bibitem{Zhong2022c}
J.~Zhong, R.~Kirby, M.~Karimi, H.~Zou, and X.~Qiu,
\newblock ``Scattering by a rigid sphere of audio sound generated by a
  parametric array loudspeaker,''
\newblock {\em J. Acoust. Soc. Am.}, vol. 151, no. 3, pp. 1615--1626, 2022.

\bibitem{Hedberg2010}
C.~Hedberg, K.~Haller, and T.~Kamakura,
\newblock ``A self-silenced sound beam,''
\newblock {\em Acoust. Phys.}, vol. 56, no. 5, pp. 637--639, 2010.

\bibitem{Zhu2023}
Y.~Zhu, W.~Ma, Z.~Kuang, M.~Wu, and J.~Yang,
\newblock ``Optimal audio beam pattern synthesis for an enhanced parametric
  array loudspeaker,''
\newblock {\em J. Acoust. Soc. Am.}, vol. 154, no. 5, pp. 3210--3222, 2023.

\bibitem{Nomura2012}
H.~Nomura, C.~M. Hedberg, and T.~Kamakura,
\newblock ``Numerical simulation of parametric sound generation and its
  application to length-limited sound beam,''
\newblock {\em Appl. Acoust.}, vol. 73, no. 12, pp. 1231--1238, 2012.

\bibitem{Nomura2025}
H.~Nomura and T.~Nakagawa,
\newblock ``Distortion reduction of length-limited sound beam generated by
  parametric acoustic array,''
\newblock {\em Appl. Acoust.}, vol. 230, pp. 110425, 2025.

\bibitem{Zhong2021}
J.~Zhong, R.~Kirby, and X.~Qiu,
\newblock ``The near field, westervelt far field, and inverse-law far field of
  the audio sound generated by parametric array loudspeakers,''
\newblock {\em J. Acoust. Soc. Am.}, vol. 149, no. 3, pp. 1524--1535, 2021.

\bibitem{Li2024}
S.-Z. Li, T.~Zhuang, J.-X. Zhong, and J.~Lu,
\newblock ``Extended king integral for modeling of parametric array
  loudspeakers with axisymmetric profiles,''
\newblock {\em J. Acoust. Soc. Am.}, vol. 156, no. 4, pp. 2189--2199, 2024.

\bibitem{Cervenka2013}
M.~Cervenka and M.~Bednarik,
\newblock ``Non-paraxial model for a parametric acoustic array,''
\newblock {\em J. Acoust. Soc. Am.}, vol. 134, no. 2, pp. 933--938, 2013.

\bibitem{Lee2009}
H.~Lee, D.~Kang, and W.~Moon,
\newblock ``A micro-machined source transducer for a parametric array in air,''
\newblock {\em J. Acoust. Soc. Am.}, vol. 125, no. 4, pp. 1879--1893, 2009.

\bibitem{Je2015}
Y.~Je, H.~Lee, K.~Been, and W.~Moon,
\newblock ``A micromachined efficient parametric array loudspeaker with a wide
  radiation frequency band,''
\newblock {\em J. Acoust. Soc. Am.}, vol. 137, no. 4, pp. 1732--1743, 2015.

\bibitem{Wygant2009}
I.~O. Wygant, M.~Kupnik, J.~C. Windsor, W.~M. Wright, M.~S. Wochner, G.~G.
  Yaralioglu, M.~F. Hamilton, and B.~T. Khuri-Yakub,
\newblock ``50 khz capacitive micromachined ultrasonic transducers for
  generation of highly directional sound with parametric arrays,''
\newblock {\em IEEE Trans. Ultrason., Ferroelectr. Freq. Control}, vol. 56, no.
  1, pp. 193--203, 2009.

\bibitem{Hwang2016}
Y.~Hwang, Y.~Je, H.~Lee, J.~Lee, C.~Lee, W.~Kim, and W.~Moon,
\newblock ``A parametric array ultrasonic ranging sensor with electrical beam
  steering capability,''
\newblock {\em Acta Acust. United Acust.}, vol. 102, no. 3, pp. 423--427, 2016.

\bibitem{Ahn2019}
H.~Ahn, K.~Been, I.-D. Kim, C.~H. Lee, and W.~Moon,
\newblock ``A critical step to using a parametric array loudspeaker in mobile
  devices,''
\newblock {\em Sensors}, vol. 19, no. 20, pp. 4449, 2019.

\bibitem{Niu2022}
X.~Niu, Z.~Liu, Y.~Meng, C.~M. Hodges, R.~P. Williams, and N.~A. Hall,
\newblock ``An air-coupled electrostatic ultrasound transducer using a mems
  microphone architecture,''
\newblock {\em J. Microelectromech. Syst.}, vol. 31, no. 5, pp. 813--819, 2022.

\end{thebibliography}

\end{document}